\documentclass[preprint,3p]{elsarticle}

\usepackage{amsmath,graphicx,amssymb}
\usepackage{epsf,psfig}

\journal{Mechanics Research Communications}

\begin{document}

\begin{frontmatter}

\title{Chaotic orbits in a 3D galactic dynamical model with a double nucleus}

\author{Nicolaos D. Caranicolas}
\author{Euaggelos E. Zotos\corref{}}

\address{Department of Physics, \\
Section of Astrophysics, Astronomy and Mechanics, \\
Aristotle University of Thessaloniki \\
GR-541 24, Thessaloniki, Greece}

\cortext[]{Corresponding author: \\
\textit{E-mail address}: evzotos@astro.auth.gr (Euaggelos E. Zotos)}

\begin{abstract}
A 3D-dynamical model is constructed for the study of motion in the central regions of a disk galaxy with a double nucleus. Using the results of the 2D-model, we find the regions of initial conditions in the $\left(x,p_x,z,p_y\right)=E_J$, $\left(y=p_z=0\right)$ phase space producing regular or chaotic orbits. The majority of stars are on chaotic orbits. All chaotic orbits come arbitrary close to one or to both nuclei. Regular orbits are those starting near the stable periodic orbits of the 2D-system with small values of $z_0$. All regular orbits circulate around only one of the two nuclei.
\end{abstract}


\end{frontmatter}

\section{Introduction}

Astronomers are not sure how common the double nucleus phenomenon is. This happens because the centers of galaxies are obscured by dust and gas and, therefore, have yield few of their secrets. Fortunately this is beginning to change with the advent of space-based observations. Galaxies with reported double nuclei are M31 (Statler et al., 1999), the  barred spiral galaxy M83 (Mast et al., 2006; Soria \& Wu, 2002), NGC 6240 (Komosca et al., 2003; Risaliti et al., 2006), NGC 3256 (Lira et al., 2007) and the dwarf elliptical galaxy VCC 128 (Debattista et al., 2006). A detailed catalog of disk galaxies with double nuclei was compiled by Cimeno et al. (2004).

On this basis, there is no doubt that the construction of a 3D dynamical model in order to study the motion in a galaxy with a double nucleus would be of interest. As far as the authors know there are few dynamical models for the study of motion in galaxies with double nuclei (see Jalali \& Rafie, 2001; Samlhus \& Sridhar, 2002).

This article can be considered as a continuation of the results presented in a recent paper (see Caranicolas \& Papadopoulos, 2009; hereafter CP). In Section 2 we present the dynamical model. In Section 3 the regular or chaotic character of orbits in the 2D model is explored. In Section 4 we study the nature of motion in the 3D model. A discussion and the conclusions of this research are presented in Section 5.

\section{The 3D-model}

The model is an extension of the model used in CP in the 3D space with an additional disk. Thus the potential of the first body is
\begin{equation}
V_1(x,y,z) = -\frac{M_d}{\left[x^2 + y^2 +\left(\alpha + \sqrt{h^2 + z^2}\right)^2\right]^{1/2}}
- \frac{M_{n1}}{\left(x^2 + y^2 + z^2 + c_{n1}^2\right)^{1/2}},
\end{equation}
where $M_d$, $M_{n1}$ is the mass of disk and nucleus 1 respectively, $\alpha$ is the disk's scale length, $h$ is the disk's scale height, while $c_{n1}$ is the scale length of nucleus 1. The second nucleus is described by the potential
\begin{equation}
V_2(x,y,z) = -\frac{M_{n2}}{\left(x^2 + y^2 + z^2 + c_{n2}^2\right)^{1/2}},
\end{equation}
where $M_{n2}$, $c_{n2}$ is the mass and the scale length of the nucleus 2 respectively. As in CP, the two bodies move in circular orbits in an inertial frame OXYZ with the origin at the center of mass of the system at a constant angular velocity $\Omega_p >0$.

Let $M_t=M_d+M_{n1}+M_{n2}$ be the total mass of the system and $R$ be the distance between the two bodies. A clockwise rotating frame Oxyz is used with axis Oz coinciding with the axis OZ and the axis Ox coinciding with the straight line joining the two bodies. In this frame, which rotates at an angular velocity $\Omega_p$, the two bodies have fixed positions $x_1,y_1=0$ and $x_2,y_2=0$. The total potential responsible for the motion of a test particle (star) with a unit mass is
\begin{equation}
\Phi(x,y,z) = \Phi_1(x,y,z) + \Phi_2(x,y,z),
\end{equation}
where
\begin{equation}
\Phi_1(x,y,z) = -\frac{M_d}{\sqrt{r_{a1}^2 + \left(\alpha + \sqrt{h^2 + z^2}\right)^2}}
- \frac{M_{n1}}{\sqrt{r_1^2 + c_{n1}^2}} - \frac{M_{n2}}{\sqrt{r_2^2 + c_{n2}^2}},
\end{equation}
\begin{equation}
\Phi_2(x,y,z) = -\frac{\Omega_p^2}{2}\left[\frac{M_{n2}}{M_t}r_{a2}^2 + \left(1 - \frac{M_{n2}}{M_t}\right)r_{a1}^2
- R^2\frac{M_{n2}}{M_t}\left(1 - \frac{M_{n2}}{M_t}\right)\right],
\end{equation}
and
\begin{equation}
r_{a1}^2 = (x - x_1)^2 + y^2, \ \ \ r_{a2}^2 = (x - x_2)^2 + y^2,
\end{equation}
\begin{equation}
r_1^2 = r_{a1}^2 + z^2, \ \ \ r_2^2 = r_{a2}^2 + z^2,
\end{equation}
with
\begin{equation}
x_1 = -\frac{M_{n2}}{M_t}R, \ \ \ x_2 = R - \frac{M_{n2}}{M_t}R.
\end{equation}

The angular frequency $\Omega_p$ is calculated as follows: The two bodies move about the center of mass of the system with angular velocities $\Omega_{p1}$, $\Omega_{p2}$ given by
\begin{equation}
\Omega_{p1} = \left[\frac{1}{x_1}\left(\frac{-dV_2(r)}{dr}\right)_{r=R}\right]^{1/2}, \ \ \
\Omega_{p2} = \left[\frac{1}{x_2}\left(\frac{dV_1(r)}{dr}\right)_{r=R}\right]^{1/2},
\end{equation}
where $r^2 = x^2 + y^2$. As the two bodies are not mass points the two angular frequencies are not equal. Nevertheless, we can make them equal by choosing properly the parameters $\alpha, h, c_{n1}, c_{n2}$ of the system. The authors must make clear that, after choosing properly the parameters the two frequencies may differ slightly, so that $\nu = (\Omega_{p1} - \Omega_{p2})/\Omega_{p1}$ is of order of $10^{-6}$ or smaller. Therefore, we consider the two angular frequencies practically equal, that is $\Omega_{p1} = \Omega_{p2} = \Omega_p$.

The equations of motion are written
\begin{equation}
\ddot{x} = - \frac{\partial \Phi}{\partial x} - 2 \Omega_p\dot{y}, \ \ \
\ddot{y} = - \frac{\partial \Phi}{\partial y} + 2 \Omega_p\dot{x}, \ \ \
\ddot{z} = - \frac{\partial \Phi}{\partial z},
\end{equation}
where the dot indicates derivative with respect to the time. The only integral of motion for the system of differential equations (10) is
\begin{equation}
J = \frac{1}{2} \left(p_x^2 + p_y^2 + p_z^2 \right) + \Phi(x,y,z) = E_J,
\end{equation}
where $p_x$, $p_y$ and $p_z$ are the momenta per unit mass conjugate to $x$, $y$ and $z$ respectively. This is the well known Jacobi integral and $E_J$ is its numerical value. In this work we use the same system of galactic units, as in CP. The unit of length is 1 kpc, the unit of  mass is 2.325 $\times 10^7$ $M_{\odot}$  and the unit of time is 0.977 $\times 10^8$ yr. The velocity unit is equal to 10 km/s, while $G$ is equal to unity.

\section{Orbits in the 2D potential}

In this section we shall study the 2D potential. Thus we set $z=p_z=0$ in (11) obtaining
\begin{equation}
J_2 = \frac{1}{2}\left(p_x^2 + p_y^2 \right) + \Phi(x,y) = E_{J2},
\end{equation}
where $E_{J2}$ is the numerical value of $J_2$. As the system is now two-dimensional, we can use the classical method of the $(x,p_x)$, $y=0, p_y>0$, Poincar\'{e} phase plane in order to explore the regular or chaotic character of motion. The results obtained from the study of the 2D system will be used in order to determine the character of orbits in the 3D system.

Our results come from the numerical integration of the equations of motion, which was done using a Bulirsh-St\"{o}er method in double precision. The accuracy of the calculations was checked by the constancy of the Jacobi integral, which was conserved up to the twelfth significant figure.

Fig. 1 shows the $(x,p_x)$ phase plane when $M_d=1100, M_{n1}=100, M_{n2}=600, R=1.5, c_{n1}=0.10, c_{n2}=0.35, \Omega_p=22.1938, \alpha=0.3053, h=0.06$, while $E_{J2}=-2010$. As one can see there is a large unified chaotic sea surrounding both nuclei. One also observes two separate regular regions near each nucleus. It is of interest to note that the regular regions are around the retrograde periodic points in each of the two nuclei. Some smaller regular regions are also present near the smaller nucleus 1 on the left. One additional feature is the presence of several small islands  near  the heavy nucleus 2 on the right. These small islands are produced by secondary resonances.

Fig. 2 is similar to Fig. 1 but when $M_d=1100, M_{n1}=300, M_{n2}=400, R=1.5, c_{n1}=0.10, c_{n2}=0.25, \Omega_p=22.6244, \alpha=0.2181, h=0.06$, while $E_{J2}=-2100$. Here again we see a large unified chaotic sea surrounding both nuclei. The two regular areas around the  retrograde periodic points of each nucleus are also present. On the other hand, a careful observer is able to see some significant differences between the two patterns. In Fig. 2 the regular area near the nucleus 1 on the left is now considerably larger than that observed in Fig. 1. Furthermore an additional small regular region has appeared near direct periodic point of the nucleus 1 on the left. Moreover the corresponding regular area near the nucleus 2 on the right has now become smaller.
\begin{figure}
\centering
\resizebox{0.6\hsize}{!}{\rotatebox{0}{\includegraphics*{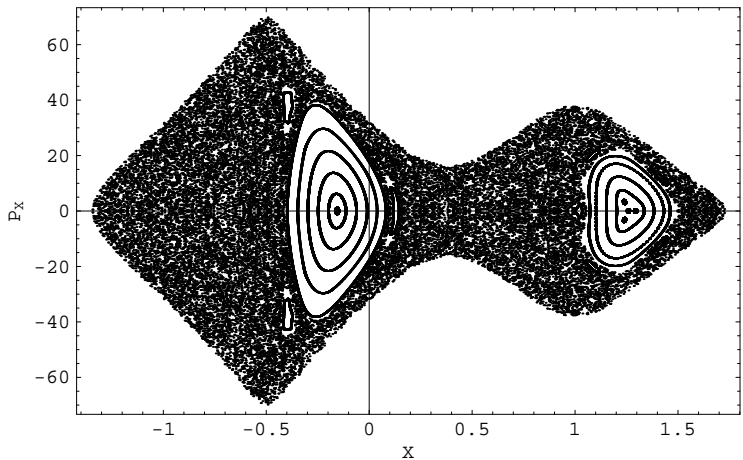}}}
\caption{The $(x,p_x)$ phase plane when $M_d=1100, M_{n1}=100, M_{n2}=600, R=1.5,$
$c_{n1}=0.10, c_{n2}=0.35, \Omega_p=22.1938, \alpha=0.3053, h=0.06$. The value of $E_{J2}$ is equal to $-2010$.}
\end{figure}
\begin{figure}
\centering
\resizebox{0.6\hsize}{!}{\rotatebox{0}{\includegraphics*{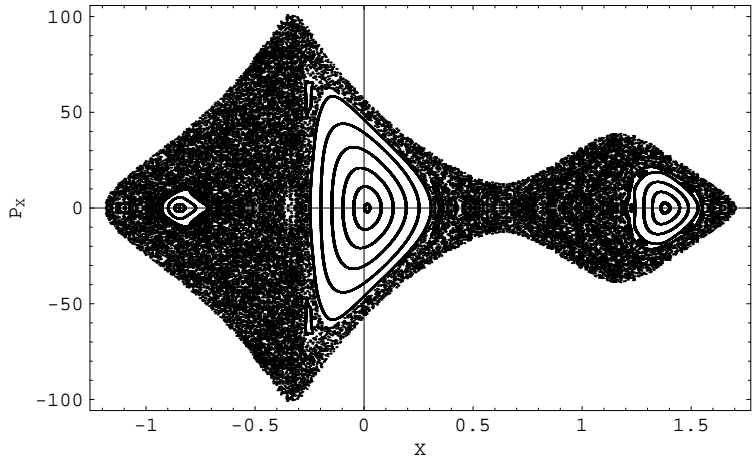}}}
\caption{Similar to Fig. 1 but when $M_d=1100, M_{n1}=300, M_{n2}=400, R=1.5,$
$c_{n1}=0.10, c_{n2}=0.25, \Omega_p=22.6244, \alpha=0.2181, h=0.06$, while $E_{J2}=-2100$.}
\end{figure}

This can be explained, if we take into account that the observed chaotic motion near each nucleus is a result not only of the force coming from the nucleus itself but also of the force coming from the other nucleus. As the mass of nucleus 2 is smaller while the mass of the disk and the distance $R$ between the two nuclei are the same, the influence from this nucleus to the disk and the nucleus 1 system is smaller. This has as a result the reduction of chaotic region and a parallel increase of the regular regions associated with the nucleus 1 and the disk. On the other hand, for similar reasons, as the mass of nucleus 1 has increased, the regular area associated with nucleus 2 has become smaller.

The numerical results suggest that there are two kinds of chaotic orbits: (i) chaotic orbits approaching  both nuclei and (ii) chaotic orbits that approach only one of the two nuclei. On the other hand, the regular orbits circulate around only one of the two nuclei.

\section{Orbits in the 3D potential}

In this section we shall study the properties of orbits in the 3D potential. In order to keep things simple we shall use our experience gained from the study of the 2D system in order to obtain a clear picture of the properties of orbits in the 3D dynamical model.

We are particularly interested to locate the initial conditions in the 3D model producing regular or chaotic orbits. A convenient way to obtain this is to start from the $(x,p_x)$ phase plane of the 2D system with the same value of the Jacobi integral used in the 2D system. Thus we take $E_J=E_{J2}$. For this purpose a large number of orbits were computed with initial conditions $(x_0,p_{x0},z_0)$, where $(x_0,p_{x0})$ is a point in the chaotic sea of Figs. 1 or 2 with all permissible values of $z_0$, and $p_{z0}=0$. Remember that, as we are on the phase plane, we have $y_0=0$, while in all cases the value of $p_{y0}$ was obtained from the Jacobi integral. All tested orbits were found to be chaotic. Therefore, one concludes that the majority of orbits in the 3D system are chaotic.

An interesting question one might ask is this. Are there any other chaotic orbits in the 3D system? In order to give an answer we have taken the sections of the 3D orbit with the plane $y=0$, when $p_y >0$. The set of the four-dimensional points $(x,p_x,z,p_z)$ was projected on the $(z,p_z)$ plane. If the projected points lie on an ``invariant curve" this suggests that the motion is regular, while, if not, this is an indication that the motion is chaotic. Fig. 3 shows such ``invariant curves" for orbits starting near the regular region on the right side of Fig. 1. In order to obtain the results shown in Fig. 3 we have taken the point $(x_0,p_{x0})=(-0.17,0.0)$ representing approximately the position of the periodic orbits in the $(x,p_x)$, $y=0, p_y>0$, phase plane and a set of values of $z_0=(0.02, 0.05, 0.08, 0.11, 0.15, 0.20, 0.25)$. Note that the numerical results indicate that, for small values of $z_0$ the motion is regular, while for larger values of $z_0$, the motion seems to be chaotic. Here we must emphasize, that the results of Fig. 3 are rather qualitative and can be considered as an indication that the transition from regularity to chaos occurs as the value of $z_0$ increases. Results not given here show a similar behavior near each regular region in Figs. 1 and 2.
\begin{figure}[!tH]
\centering
\resizebox{0.6\hsize}{!}{\rotatebox{0}{\includegraphics*{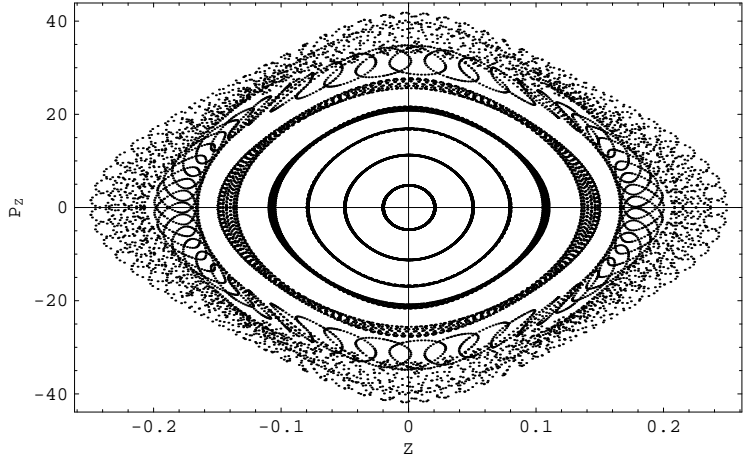}}}
\caption{Projection of the sections of the 3D orbit with the plane $y=0$, when $p_y >0$. The set of the four dimensional points $(x,p_x,z,p_z)$ is projected on the $(z,p_z)$ plane. }
\end{figure}

In order to estimate the degree of chaos in the 2D and the 3D system we have calculated the maximum LCE (Lyapunov Characteristic Exponent) (see Lichtenberg \& Lieberman, 1992) for a large number of chaotic orbits. Each LCE was calculated for a time period of $10^4$ time units. The LCE for the 2D orbits, starting in the chaotic sea of Fig. 1, was found in the range $5.0-5.2$, while in the chaotic sea of Fig. 2 it was found in the range $6.0-6.2$. The LCE for chaotic orbits in the 3D system with initial conditions $(x_0, p_{x0}, z_0)$ with $(x_0,p_{x0})$ in the chaotic sea of the Fig. 1 was found in the range $3.2-3.4$. The corresponding values of LCE for Fig.2 were found in the range $3.9-4.1$.
\begin{figure*}[!tH]
\centering
\resizebox{0.8\hsize}{!}{\rotatebox{0}{\includegraphics*{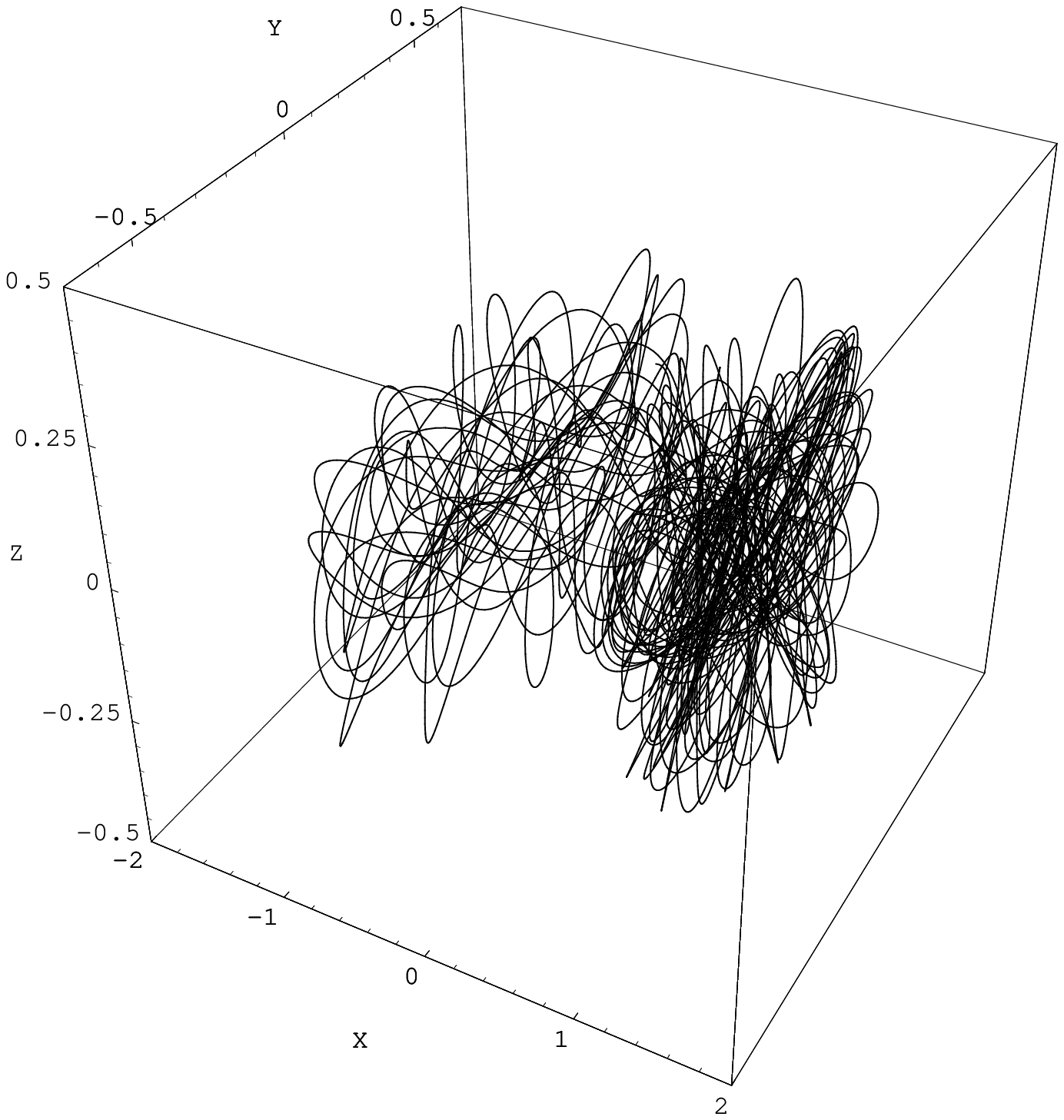}}\hspace{1cm}
                         \rotatebox{0}{\includegraphics*{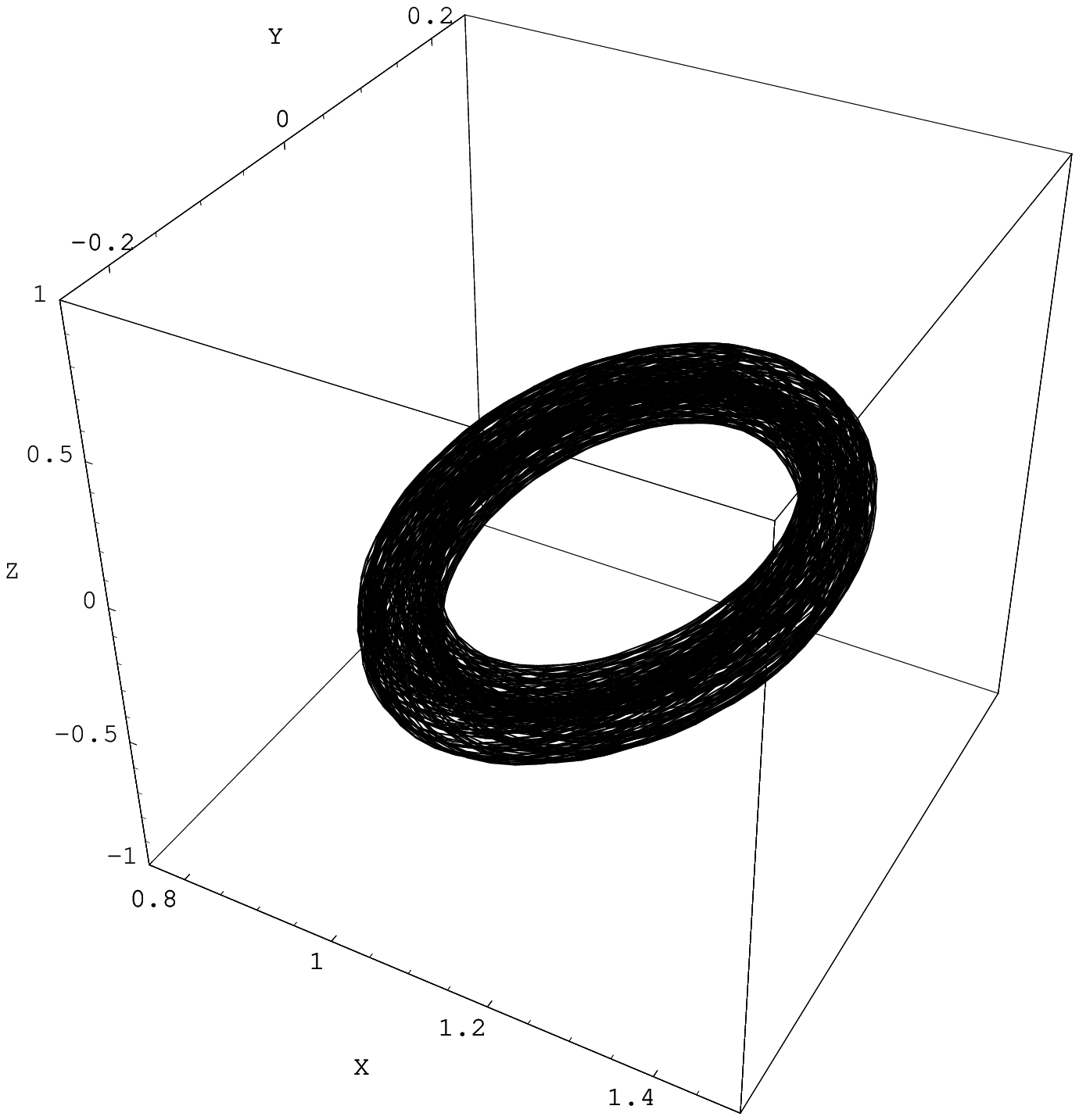}}}
\vskip 0.1cm
\caption{(a-b): (a-left) A chaotic orbit approaching both nuclei. The initial conditions are $x_0=0.8, y_0=p_{x0}=0, z_0=0.1, p_{z0}=0$. The values of the other parameters are as in Fig. 1, while $E_J=-2010$. (b-right) A regular orbit circulating around nucleus 1 with initial conditions $x_0=1.4, y_0=p_{x0}=0, z_0=0.15, p_{z0}=0$. The values of the other parameters are as in Fig. 2 and $E_J=-2100$.}
\end{figure*}
\begin{figure*}[!tH]
\centering
\resizebox{0.8\hsize}{!}{\rotatebox{0}{\includegraphics*{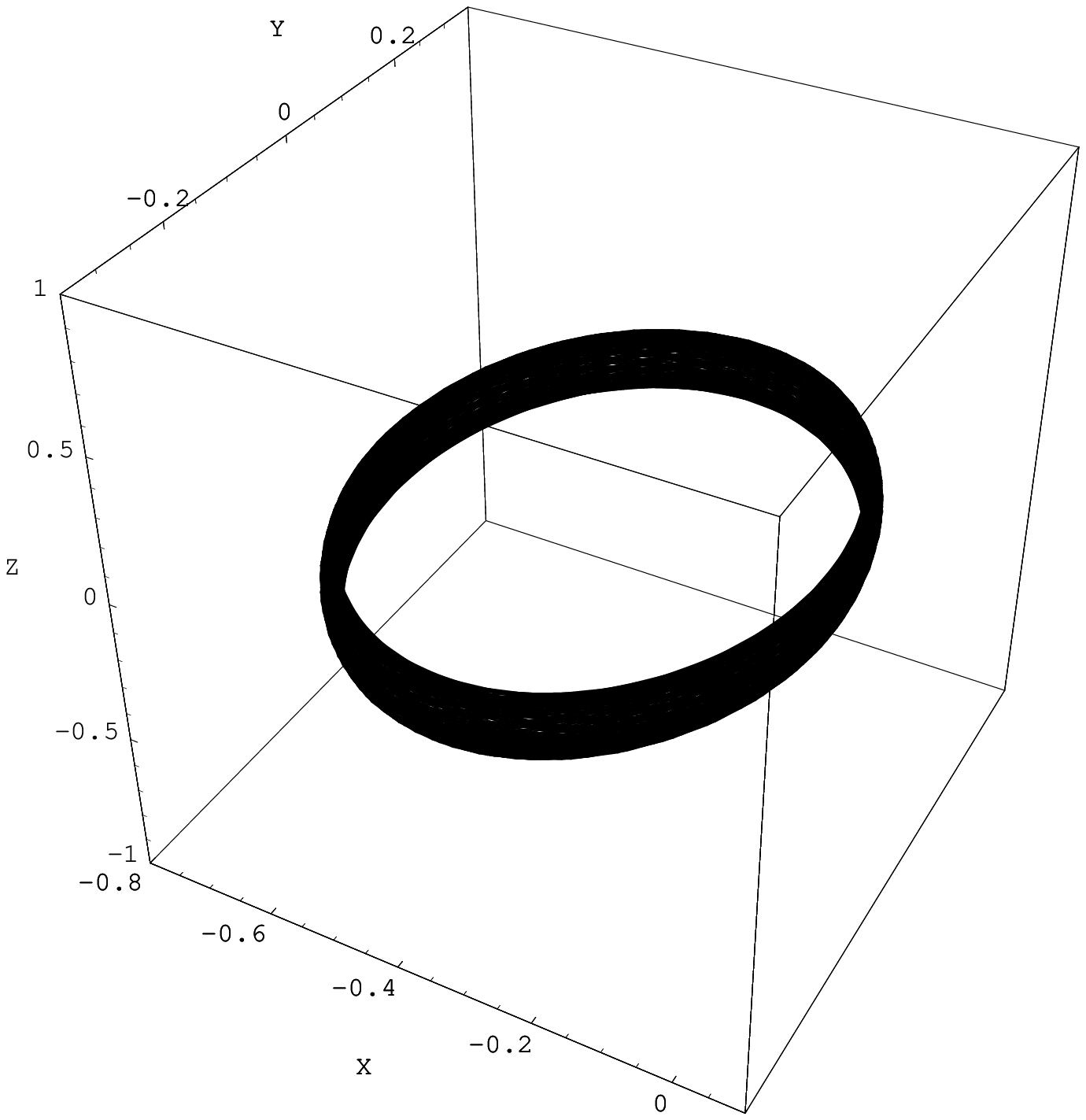}}\hspace{1cm}
                         \rotatebox{0}{\includegraphics*{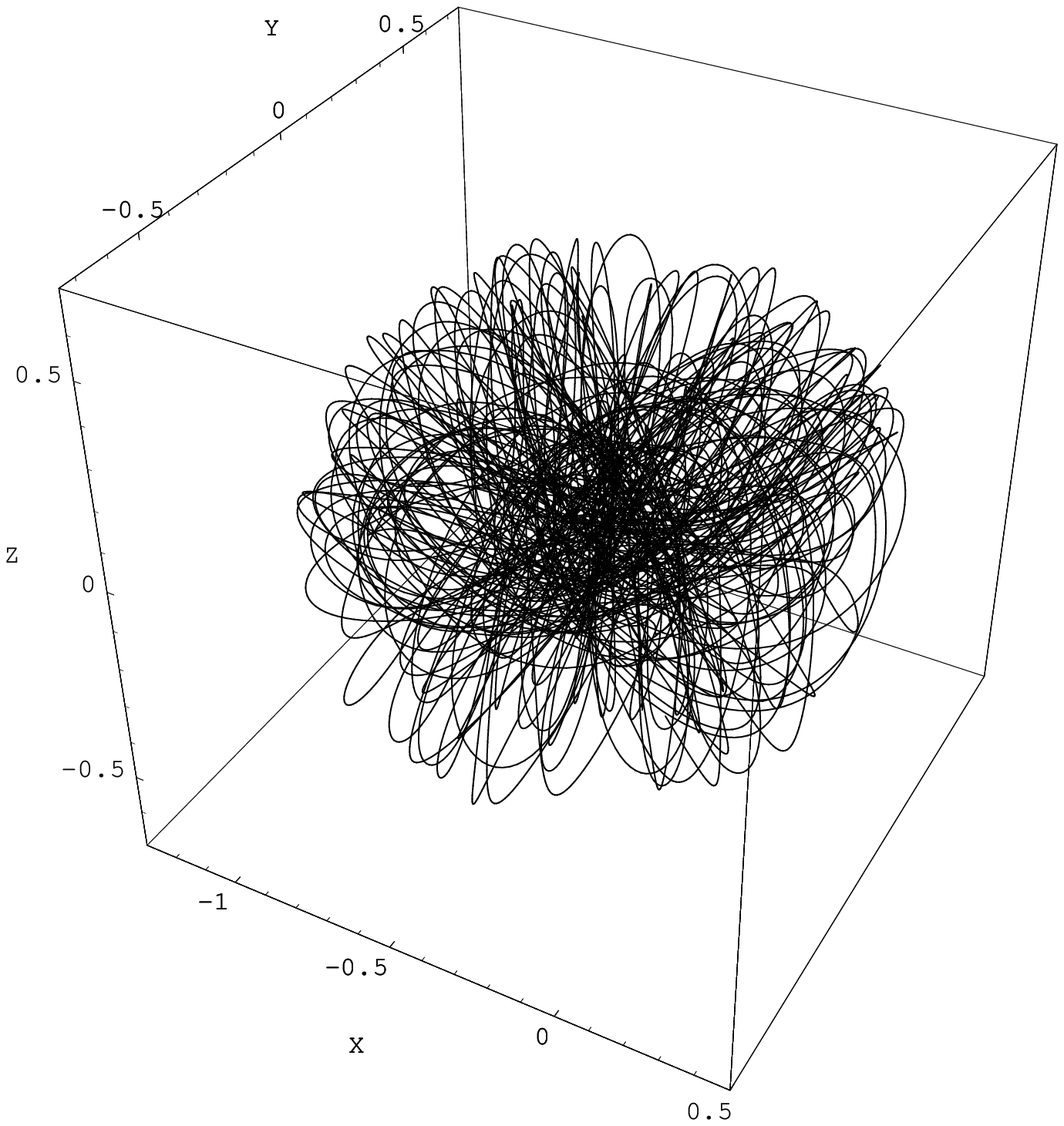}}}
\vskip 0.1cm
\caption{(a-b): (a-left) A regular orbit and (b-right) a chaotic orbit. The two orbits differ in initial conditions only in the value of $z_0$. See text for details.}
\end{figure*}
\begin{figure*}[!tH]
\centering
\resizebox{0.8\hsize}{!}{\rotatebox{0}{\includegraphics*{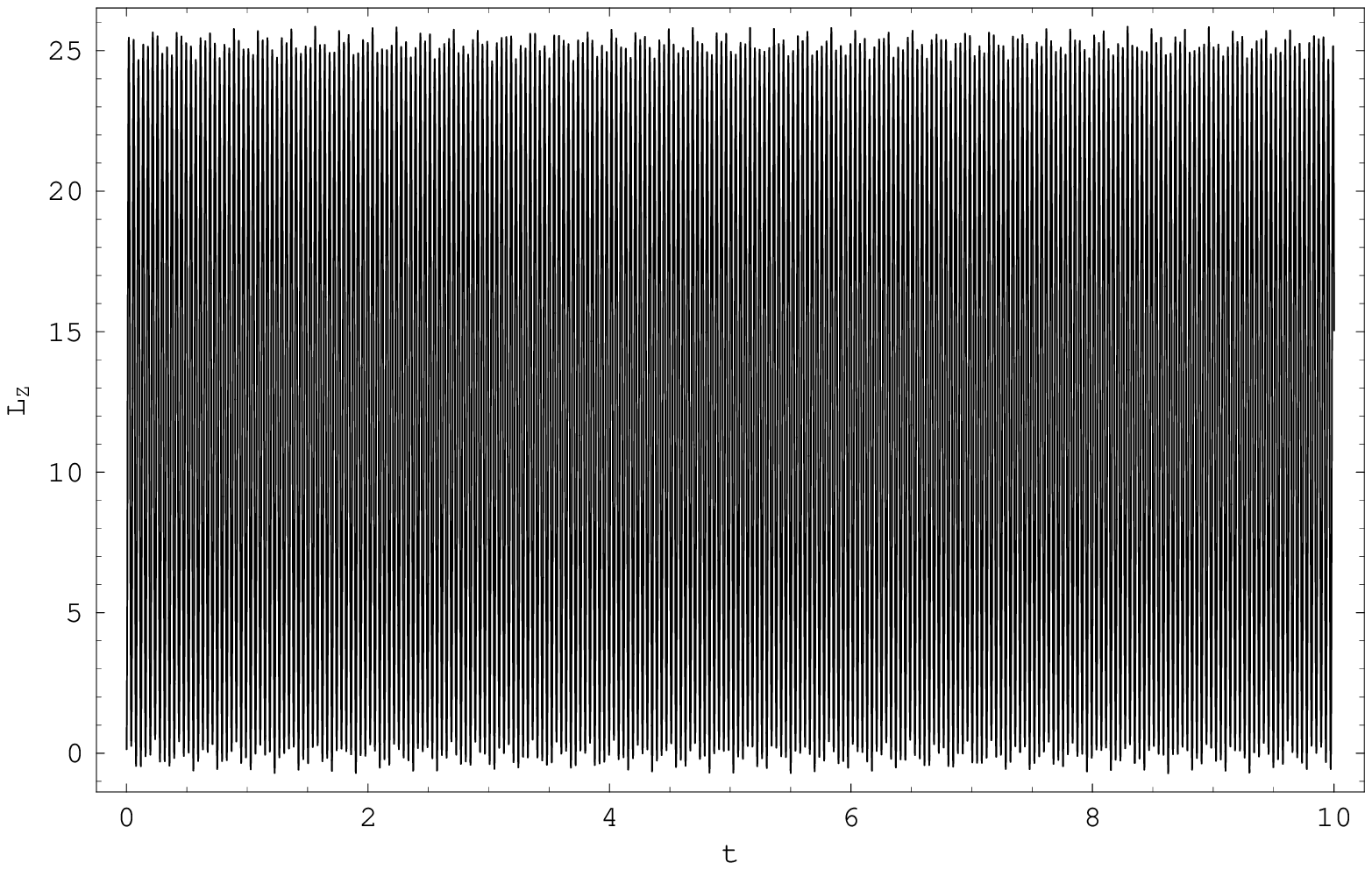}}\hspace{1cm}
                         \rotatebox{0}{\includegraphics*{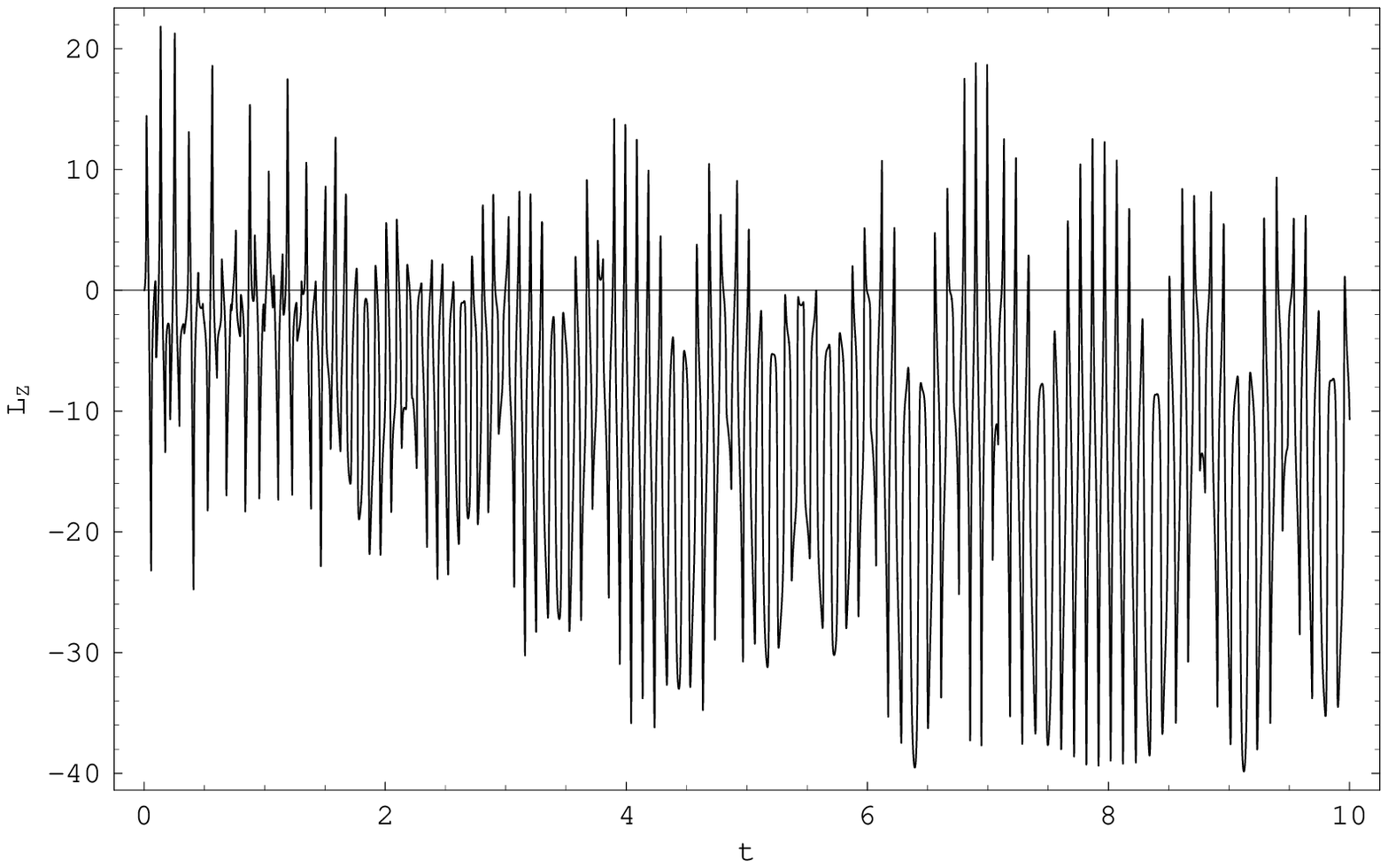}}}
\vskip 0.1cm
\caption{(a-b): (a-left) Evolution of $L_z$ with the time for the regular orbit of Fig. 5a and (b-right) for the chaotic orbit of Fig. 5b.}
\end{figure*}

Fig. 4a shows a chaotic orbit with conditions $x_0=0.8, y_0=p_{x0}=0, z_0=0.1, p_{z0}=0$. The value of $p_{y0}$ is always found from the Jacobi integral. The values of the other parameters are as in Fig. 1, while $E_J=-2010$. Note that the orbit goes arbitrary close to both nuclei. It is interesting to observe that near the more massive nucleus the orbit is deflected to more higher values of $z$, while near the less massive nucleus the star stays close to the disk. Fig. 4b shows a regular orbit circulating around nucleus 1. The initial conditions are $x_0=1.4, y_0=p_{x0}=0, z_0=0.15, p_{z0}=0$. The values of the other parameters are as in Fig. 2, while $E_J=-2100$.

Fig. 5a shows a quasi-periodic orbit starting near the retrograde periodic point, which is close to nucleus 1. The initial conditions are $x_0=0, y_0=p_{x0}=0, z_0=0.1, p_{z0}=0$. The values of the other parameters are as in Fig. 2, while $E_J=-2100$. Fig. 5b shows an orbit with the same initial conditions, the same value of the Jacobi integral and the same values of the parameters as in Fig. 5a but when $z_0=0.5$. The orbit has now become chaotic and goes arbitrarily close to the nucleus 1. This orbit shows, from another point of view, that 3D orbits starting near the stable periodic points of the 2D system are regular only for small values of $z_0$.

Here the physical parameter playing an important role is the $L_z$ component of the angular momentum. From our previous experience we know that low angular momentum stars, on approaching a dense nucleus are scattered off the galactic plane displaying chaotic motion (Caranicolas \& Innanen, 1991; Caranicolas \& Papadopoulos, 2003). Of course here in 3D space things are more complicated than in an axially symmetric dynamical model, where $L_z$ is conserved (see Caranicolas \& Innanen, 1991). As $L_z$, is not conserved, we can compute numerically the mean value $< L_z >$ of $L_z$ using the formula
\begin{equation}
< L_z > = \frac{1}{n}\displaystyle\sum_{i}^{n}L_{zi}.
\end{equation}

Our numerical calculations suggest that the chaotic orbits have low values of $< L_z >$, while regular orbits have high values of $< L_z >$. Fig. 6a shows the evolution of $L_z$ with the time for the regular orbit of Fig. 5a. Here we find a value of $< L_z >=12.5$. Fig. 6b is similar to Fig. 6a but for the chaotic orbit of Fig. 5b. Here $< L_z >=-10.3$ The value of $n$ in both cases was $10^4$.

\section{Discussion and conclusions}

Observation data show that a small fraction of active galaxies have double nuclei (Eracleus \& Halpen, 2003; Xinwu \& Ting-Gui, 2006). It was this reason that motivated our construction of a 3D model in order to study the motion in a disk galaxy hosting a binary nucleus. It was found that the majority of orbits in the 2D system were chaotic. Two kinds of chaotic orbits were observed: (i) Chaotic orbits that approach both nuclei and (ii) Chaotic orbits that approach only one of the nuclei. The regular regions are confined mainly around  the retrograde periodic points in both nuclei. All regular orbits go around nucleus 1 or nucleus 2 but not both. It was also found that the total velocity near each nucleus attains high values. The value of velocity depends on the mass of the nucleus and the value of its scale length. Regular motion corresponds to low velocities while chaotic motion is characterized by high velocities.

In order to understand the behavior of orbits in the 3D system we have used our knowledge from the 2D system. Of particular interest was the determination of the region of initial conditions in the $(x,p_x,z,p_y)=E_J$, $(y=p_z=0)$ phase space that produces regular or chaotic 3D orbits. As $p_{y0}$ was found always from the Jacobi integral we have used the same value of $E_J$, as in the 2D system and took initial condition $(x_0,p_{x0},z_0)$ such as $(x_0,p_{x0})$ lies in the chaotic region of the 2D system. It was found that the motion was chaotic for all permissible values of $z_0$. In the case when $(x_0,p_{x0})$ was inside a regular region, the corresponding 3D orbit was regular only for small values of $z_0$, while for larger values of $z_0$ the orbit was chaotic. The particular values of $z_0$ were different for each regular region of the 2D system. We did not feel that it was necessary to try to define the particular values of $z_0$ for each case.

An important role is played by the $L_z$ component of the test particle's angular momentum. It was found that the values of $< L_z >$ for regular orbits are larger, than those corresponding to chaotic orbits. Thus, the $L_z$ component of the angular momentum is a significant parameter connected with the regular or chaotic character of orbits in both 2D and 3D galactic models.

In order to estimate the degree of chaos in the 2D as well as in the 3D potentials, we have computed the maximum Lyapunov Characteristic Exponent (LCE) for a large number of orbits for a time period of $10^4$ time units. The numerical results indicate that the degree of chaos in 3D double nucleus systems is smaller than in similar 2D systems.

\section*{Acknowledgement}

\textit{The authors would like to thank an anonymous referee for his useful suggestions and comments.}

\end{document}